\documentclass[12pt,preprint]{aastex}
\def\h1{$^1$H}
\def\o16{$^{16}$O}
\def\c12{$^{12}$C}
\def\n14{$^{14}$N}

\def\he4{$^4$He}

\everymath{\rm}
\everydisplay{\sf}
\received{}
\revised{}
\accepted{}
\cpright{}{}\journalid{}{}
\articleid{}{}
\paperid{}
\shortauthors{Henry et al.}
\shorttitle{DdDm-1}
\begin{document}
\title{A MULTIWAVELENGTH ANALYSIS OF THE HALO PLANETARY NEBULA DdDm-1\altaffilmark{1,2}} 

\author{R.B.C. Henry\altaffilmark{3}}

\affil{H.L. Dodge Department of Physics \& Astronomy, University of Oklahoma; rhenry@ou.edu}

\author{K.B. Kwitter\altaffilmark{3}}

\affil{Department of Astronomy, Williams College; kkwitter@williams.edu}

\author{R.J. Dufour}

\affil{Department of Physics \& Astronomy, Rice University; rjd@rice.edu}
 
\and 
 
\author{J.N. Skinner}
\affil{Department of Physics and Astronomy, Dartmouth College; Julie.N.Skinner@Dartmouth.edu}

\altaffiltext{1}{This work is based in part on observations made with the Spitzer Space Telescope, which is operated by the Jet Propulsion Laboratory, California Institute of Technology under a contract with NASA. Support for this work was provided by NASA through an award issued by JPL/Caltech.}
\altaffiltext{2}{Based on observations made with the NASA/ESA Hubble Space Telescope, obtained from the data archive at the Space Telescope Institute. STScI is operated by the association of Universities for Research in Astronomy, Inc. under the NASA contract  NAS 5-26555.}
\altaffiltext{3}{Visiting Astronomer, Kitt Peak National Observatory, National Optical Astronomy Observatory, which is operated by the Association of Universities for Research in Astronomy, Inc. (AURA) under cooperative agreement with the National Science Foundation.}

\begin{abstract}

We present new HST optical imagery as well as new UV and IR spectroscopic data obtained with the Hubble and Spitzer Space Telescopes, respectively, of the halo planetary nebula DdDm-1. For the first time we present a resolved image of this object which indicates that the morphology of DdDm-1 can be described as two orthogonal elliptical components in the central part surrounded by an extended halo.
The extent of the emission is somewhat larger than was previously reported in the literature. We combine the spectral data with our own previously published optical measurements to derive nebular abundances of He, C, N, O, Ne, Si, S, Cl, Ar, and Fe. Our abundance determinations include the use of the newly developed program ELSA for obtaining abundances directly from emission line strengths along with detailed photoionization models to render a robust set of abundances for this object. The metallicity, as gauged by oxygen, is found to be 0.46~dex below the solar value, confirming DdDm-1's status as a halo PN. In addition, we find that Si and Fe are markedly underabundant, suggesting their depletion onto dust. The very low (but uncertain) C/O ratio suggests that the chemistry of the nebula should be consistent with an oxygen-rich environment. We find that the sulfur abundance of DdDm-1 is only slightly below the value expected based upon the normal lockstep behavior between S and O observed in H~II regions and blue compact galaxies. The central star effective temperature and luminosity are estimated to be 55,000~K and 1000~L$_{\odot}$, respectively, implying an initial progenitor mass of $<$1~M$_{\odot}$. Finally, we report on a new radial velocity determination from echelle observations.

\end{abstract}


\clearpage

\section{INTRODUCTION}

Planetary nebulae (PNe) have long served as useful tracers of galactic interstellar abundances of elements such as O, Ne, S, Cl, and Ar, alpha elements whose abundances are expected to evolve in lockstep due to their common synthesis sites in massive stars. In a recent study of S, Cl, and Ar abundances in over 80 type~II PNe located mainly in the northern Galactic hemisphere, \citet{HKB04} discovered that while Ne and O abundances track each other closely, S abundances in a large fraction of objects fall significantly below the values expected for their O abundances and exhibit a large amount of scatter as well. These authors also showed that the same pattern is present in data in the large southern survey by \citet{KB94}. 

\citet{HKB04} suggested that this tendency of PNe to display a S deficit, a situation they dubbed the ``sulfur anomaly'', was probably due to the underestimation for many objects of the ionization correction factor used to account for unobserved ions of S when determining the total gas-phase elemental abundance from the optically observable ions of S$^+$ and S$^{+2}$. However, measurements of S$^{+3}$ abundances by \citet{D03} and the results from numerous ISO projects summarized in \citet{PBS06} using the [S~IV] 10.5$\mu$m line appear to rule out this explanation, although it is probably too early to draw definite conclusions. Instead, \citet{PBS06} suggested that the sulfur anomaly could result from the depletion of S onto dust due to the formation of sulfides such as MgS and FeS. Sulfide formation is expected to occur most readily in carbon rich environments, i.e., in objects where C/O $>$ 1.

The goal of the project described in this paper is to determine and study the gas phase abundances of the halo PN DdDm-1, an object for which we present new optical imagery as well as spectroscopic measurements of important UV and IR emission lines. DdDm-1 was chosen primarily because of the large amount of data both new (presented here) and previously published for this object.  Of the many catalogued PNe in our Galaxy, only about 12 are believed to be located in the halo.  Abundance studies of these objects can be used to determine the chemical composition of the halo and can provide additional insight into the evolution of this region of the Milky Way. These objects have consistently been shown to possess sub-solar levels of metals [\citet{TPRP81}, \citet{PTPR92}, \citet{HHM97}, \citet{D03}]. DdDm-1 has been included in many large abundance surveys but has only been studied in detail in a few instances, e.g. \citet{BC84}, \citet{D03}, and \citet{WCP05}. Here we focus on it exclusively.

In this paper we present new optical HST imagery, new UV data taken with the FOS on HST, and new IR data taken with the IRS on the Spitzer Space Telescope. The imagery allows us for the first time to resolve the nebula and study its morphology. Then combining the spectral data with our own previously published optical data, we compute the abundances of He, C, N, O, Ne, Si, S, Cl, Ar, and Fe using both empirical and numerical methods to ensure robust results. For the empirical method we employ the program ELSA \citep{J07}; the program Cloudy \citep{Fer98} is used for the numerical method. The numerical approach also allows us to infer the effective temperature and luminosity of the central star. Since the Spitzer data allow us to determine the abundance of S$^{+3}$, the primary unobservable ion when only optical measurements are available, we evaluate the accuracy of the sulfur ionization correction factor for DdDm-1. With a new sulfur abundance in hand, we then assess the S deficit of DdDm-1, as well as check on the consistency of the deficit's magnitude with the value of C/O derived from our new UV measurements. Finally, we present new echelle data for DdDm-1 and determine its radial velocity.

The paper is organized as follows. In Section 2, we discuss our observations; \textsection 3 contains a description of our procedure for computing abundances; in \textsection 4, we discuss our results and we give a summary and conclusions in \textsection 5.

\section{OBSERVATIONS}

\subsection{Optical Imagining}

\begin{figure}
   \includegraphics[width=4in,angle=0]{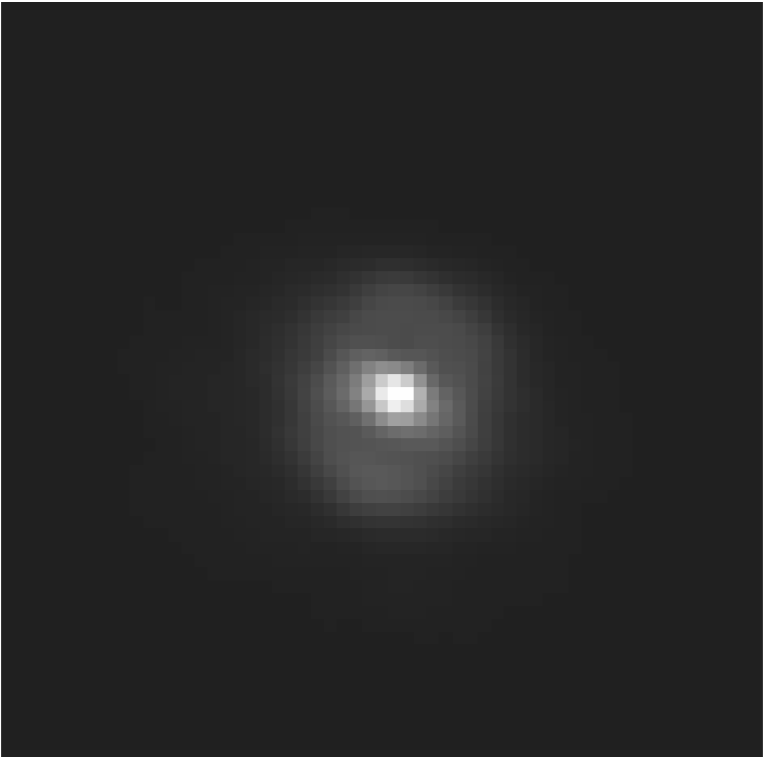} 
   \caption{Grey-scaled surface brightness image of DdDm-1 from an archival HST WFPC1 F675W 40 sec exposure taken in
1993.  North is up and east is to the left, with image dimensions of 3$\arcsec$ x 3$\arcsec$. The image quality has been partly restored using Lucy-Richardson techniques
based on a theoretical PSF for the camera/filter used. Details of the processing and morphology are given in the text.}
   \label{image1}
\end{figure}

\begin{figure}
   \includegraphics[width=4in,angle=0]{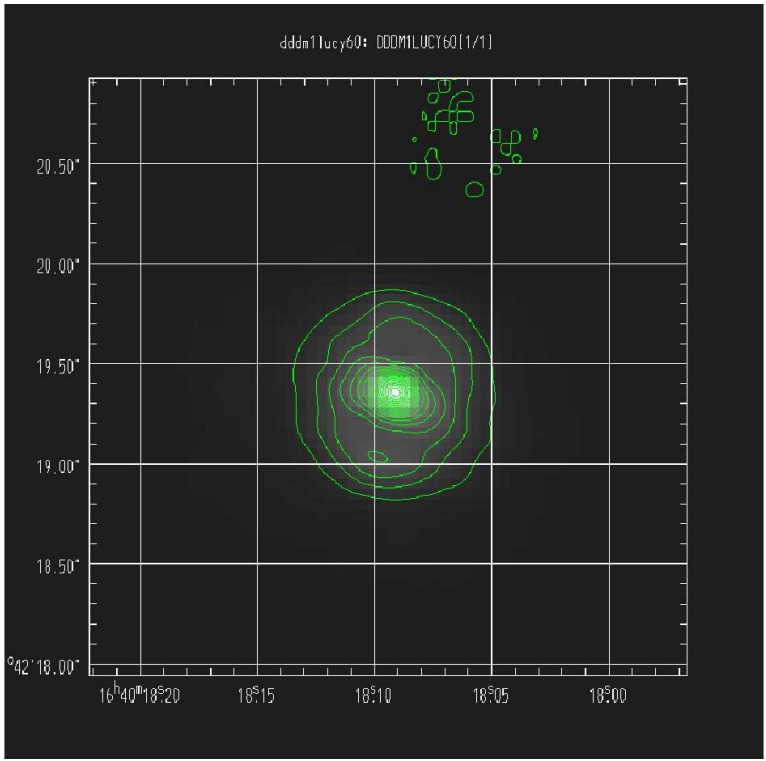} 
   \caption{Same image as in Fig. \ref{image1} but now overlaid with contours of surface brightness and the FK5
coordinate system.}
   \label{image2}
\end{figure}

In Figure \ref{image1} we present what we believe to be the first resolved image of DdDm-1 in the literature taken
from a 1993 WFPC1 image made with the Hubble Space Telescope. [Recently, \citet{WCP05} obtained an image of DdDm-1 which, according to the authors, was not well-resolved.]  The nebula was exposed
for 40 seconds through the F675W filter, which covers H$\alpha$ and nearby emission
lines. In our analysis we used the processed WPC1 image from STScI (W1j00201T), which was corrected for CCD
 artifacts and noise.  Cosmic-rays were removed using
a combination of a median filter and direct inspection. A 60$\times$60 
array ($6\arcsec\times6\arcsec$) centered on the nebula was extracted for our analysis. 
Then, the image was rebinned
into a 120$\times$120 pixel array using a cubic spline interpolation, to 
give $0.05\arcsec$ pixels.  A theoretical point-spread-function was generated
using the STScI {\it Tiny Tim}  software appropriate to the location of 
the nebula
on the WFC1, the filter, and the observation date.  This PSF was also
rebinned into 0.05$\arcsec$ pixels as per the nebula image.  The STSDAS ``Lucy"
software in the restoration package was then used to deconvolve the DdDm-1 
image with this PSF.  Our results appeared best for about 35 iterations.  
We then rotated the images by 105$^{\circ}$ counterclockwise to
align them with the equatorial coordinate system. The final image is $3\arcsec\times3\arcsec$, with north up and east to the left.
Figure \ref{image2} shows the same image as the one in Fig.~\ref{image1} but with contours added. The latter were generated using the {\it disconlab} software in IRAF.

From the deconvolved image and contoured overlay in Figs.~\ref{image1} and \ref{image2}, the structure of
DdDm-1 is seen to be elliptical with two central components possibly surrounded
by an extended (and nearly circular) halo.  Evidence for a central star is also
seen in the center of the brighter inner ellipse, which has a major axis of
$\sim$0.50$\arcsec$ along a PA = 65$^o$ and minor axis of $\sim$0.35$\arcsec$.  This is surrounded
by a fainter outer elliptical nebula with a major axis of $\sim$1.1$\arcsec$ and minor
axis $\sim$0.95$\arcsec$ which is extended approximately orthogonal to the major axis of
the inner ellipse.  These are somewhat larger dimensions for DdDm-1 than
previously reported in \citet{Acker92}.  Indeed, a logarithmic stretch of our
processed image suggests the halo of DdDm-1 may have a diameter of upwards of
$\sim$1.75$\arcsec$ (but difficult to define due to the HST spherical aberration scattered
light problems at that time).

\subsection{Spectrophotometry}

Here we report spectroscopically observed emission line fluxes in the UV, optical, and IR spectral regions. The previously unpublished UV observations were obtained with the FOS on HST, while the optical data were measured with the Goldcam on the 2.1m telescope at KPNO and reported by \citet{KH98}. We also present new IR measurements obtained with the Spitzer Space Telescope with the IRS. Finally, we report on radial velocity measurements obtained with the echelle spectrograph on the 4m telescope at KPNO. These sets of observations are described separately in detail below. 

Our complete list of emission line measurements is presented in Table~\ref{fluxes}, where the first column lists the line wavelength and identification, the second column contains the relevant f value of the reddening function, and the third and fourth columns list the raw and dereddened fluxes, respectively. Columns~5 and 6 list two sets of model-predicted line strengths pertaining to the discussion in {\S}3.2. Lines identified with bold-faced type in column~1 are used in the initial abundance analysis described in {\S\ }3.1. Because the angular size of DdDm-1 is smaller than the size of all slits employed in the observations, we are confident that the entire flux was observed in each line within each spectral range. Thus, no adjustments were necessary in order to place all line strengths on the same scale. In the following subsections we describe individually the observations obtained within the three spectral regions. Finally, in performing the dereddening calculations we used the reddening functions of \citet{S79} for the UV, \citet{SM79} for the optical, and \citet{I05} and \citet{RL85} shortward and longward of 8~$\mu$m in the IR, respectively.

\subsubsection{Ultraviolet Data}

\begin{figure}
   \centering
   \includegraphics[width=6in,angle=270]{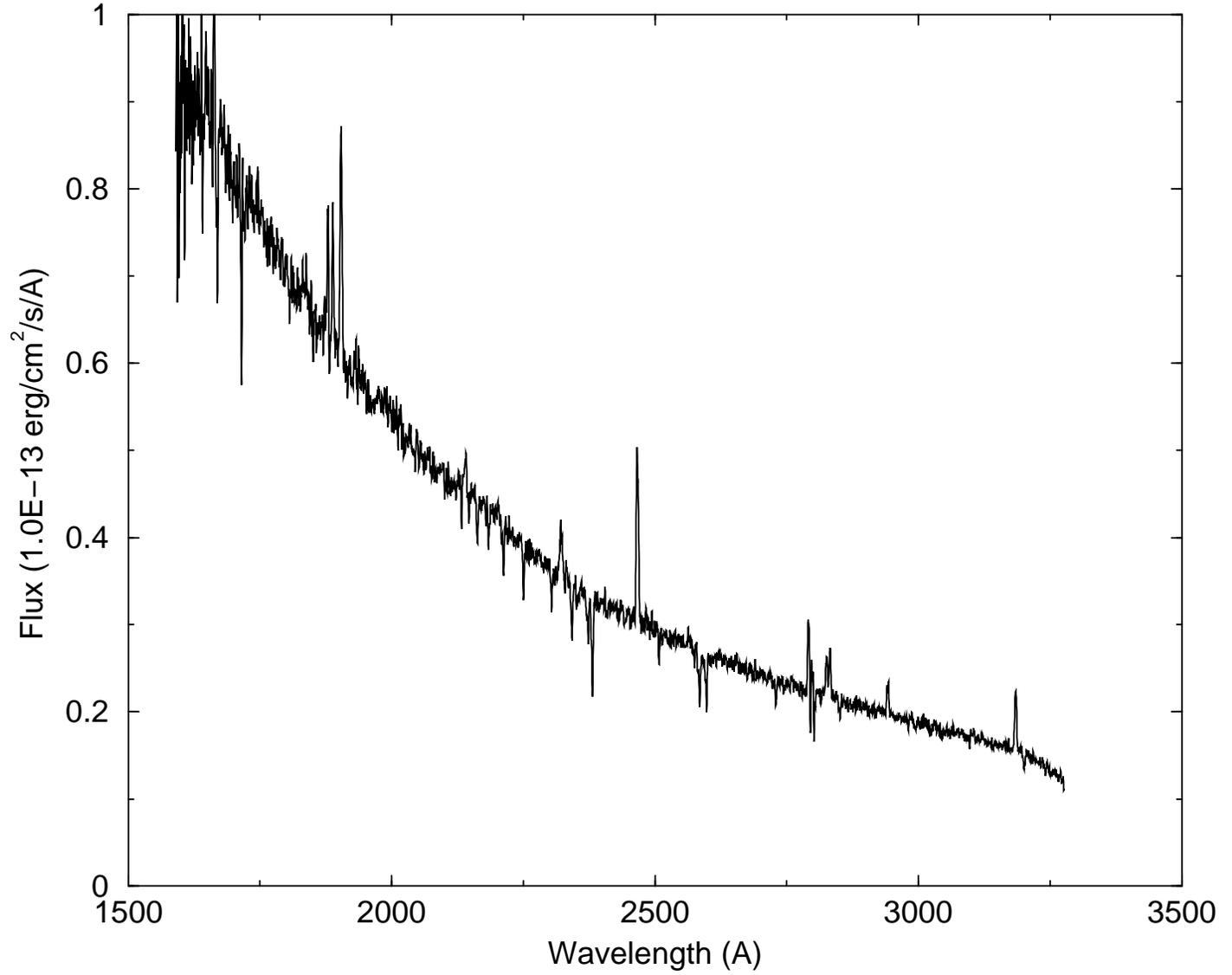} 
   \caption{Plot of the ultraviolet spectrum of DdDm-1 from archival HST FOS data.  The original spectra
(with the G190H and G270H gratings) have been smoothed by a 3-point boxcar.}
   \label{uv}
\end{figure}

DdDm-1 was observed with the Faint Object Spectrograph on the Hubble Space Telescope during
1995 October 5 as part of the Cycle 5 program GO6031.  Five ``H-series'' gratings were 
used, covering the spectral range 1087-6817\AA.  The observations were made through the
0.9 arcsec circular aperture (post-COSTAR) with a peak-up centering the central star in the
aperture.  Figure~\ref{uv} shows the ultraviolet spectrum from 1700\AA~to 3250\AA~obtained by 
splicing together the archival spectra taken with the G190H(1140 sec), and G270H(480 sec) gratings. 
Not shown is the G130H(2270 sec) spectrum which exhibits only a strong continuum and no 
obvious emission lines.  Cospatial optical wavelength spectra were also obtained with the 
G400H(90 sec) and G570H(60 sec) gratings, which permitted scaling of the UV lines to the H I
Balmer lines, enabling an accurate tie-in between the UV spectra with the ground-based optical
and Spitzer IR spectra in this study. All of the FOS spectra that were analyzed have been recalibrated by the POA-CALFOS pipeline developed by the ST-ECF in 2002 \citep{Alexov02}.

The strongest emission lines in the far-UV are dielectronic 
recombination pairs of O~III] $\lambda$1663, and Si~III] $\lambda\lambda$1882,92
(resolved and unusually strong relative to C~III]).  In the mid-UV, the dominant emission lines
are the C~II] $\lambda\lambda$2325 multiplet, [O~II] $\lambda$2470, and Mg~~II $\lambda\lambda$2795,
2803.  Table~\ref{fluxes} gives the measured strengths of the important UV abundance diagnostic lines measured
from these spectra.  Fluxes of the UV lines were measured from gaussian fits to their profiles.  We further note that
the errors in the UV line strengths are purely statistical (the square root of their FWHM times the
rms fluctuations of the nearby continuum) and do not include possible errors in the FOS calibrations or
extinction.

\subsubsection{Ground-based Optical Data}

\begin{figure}
   \centering
   \includegraphics[width=6in,angle=270]{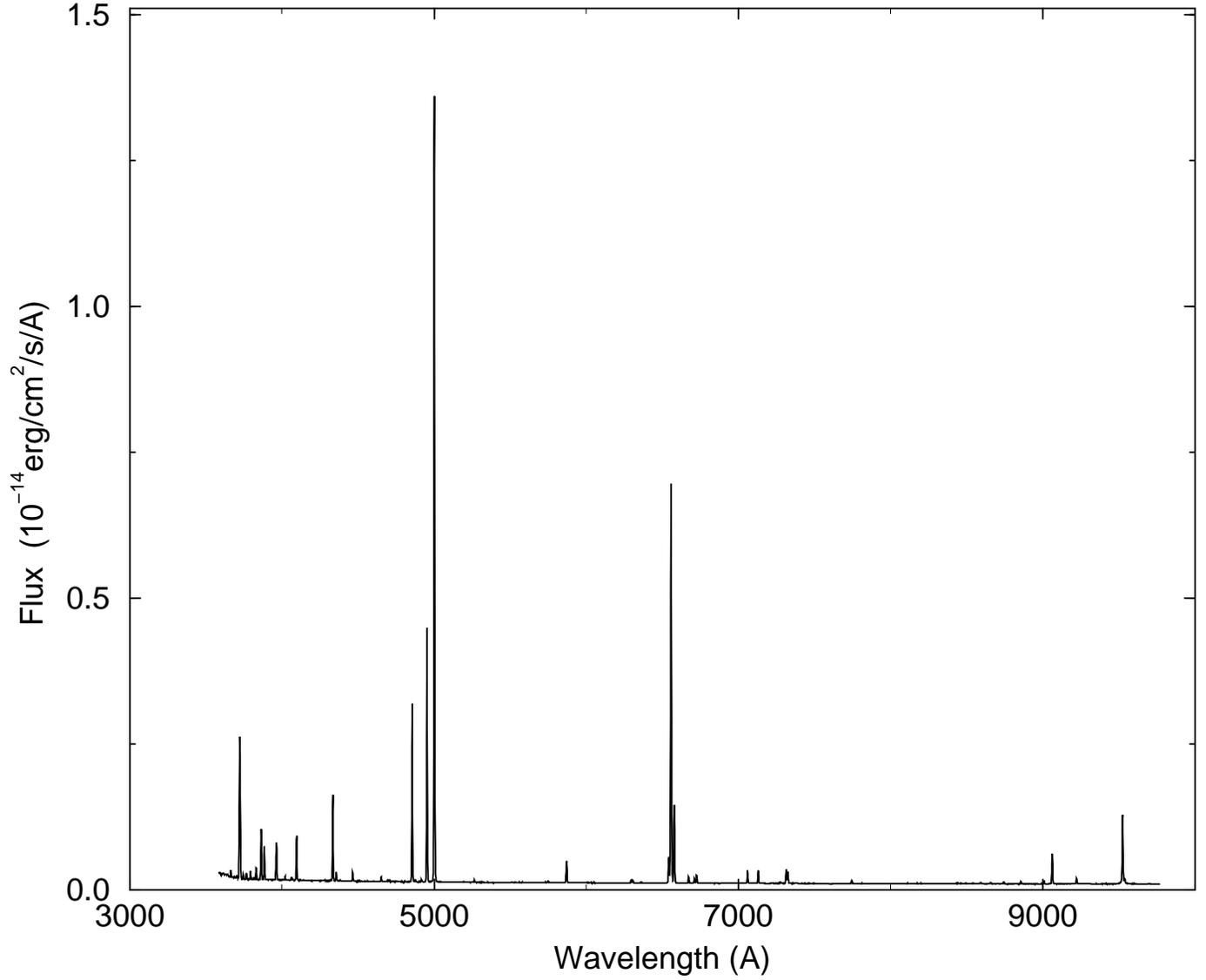} 
   \caption{Merged spectrum of DdDm-1 from KPNO observations. Note the coverage from [O~II] $\lambda$3727 to [S~III] $\lambda\lambda$9069,9532.}
   \label{optical}
\end{figure}

The optical data were obtained at Kitt Peak National Observatory in May 1996, using the 2.1m telescope and Goldcam CCD spectrograph. The spectral range from 3700-9600~\AA~was covered in two parts, with overlap from 5700-6800~\AA. The total blue integration time was 600 s and the red was 2400 s. The data were reduced with standard IRAF\footnote{IRAF is distributed by the National Optical Astronomy Observatories, which is operated by the Association of Universities for Research in Astronomy, Inc. (AURA) under cooperative agreement with the National Science Foundation.} routines.  Further details of the observations and reductions can be found in \citet{KH98}.  The merged spectrum from KPNO is shown in Fig.~\ref{optical}. We point out that our measurement of the flux in H$\beta$ agrees closely with the value reported by \citet{D03}.

\subsubsection{Infrared Data}
DdDm-1 was observed with the Infrared Spectrograph (IRS)\footnote{The IRS was a collaborative venture between Cornell University and Ball Aerospace Corporation funded by NASA through the Jet Propulsion Laboratory and Ames Research Center.} \citep{H04} on the Spitzer Space Telescope in June 2006. We used the Short Low (SL 1 and SL2), the Short-High (SH) and the Long-High (LH) modules, giving coverage from 5.2-37.2 $\mu$m. Details of the observations are given in Table~\ref{coverage}.

Spectra were extracted using SPICE, a Java tool available from the Spitzer Science Center website. Since DdDm-1 has an angular diameter of 0.6" \citep{Acker92}, i.e. smaller than the spatial resolution of all the IRS modules, it was extracted as a point source, and we presume that we have detected all of the nebular flux. This is confirmed by the measured fluxes relative to H$\beta$ of the strongest observed H transitions (9-7 at 11.3 $\mu$m and 7-6 at 12.4 $\mu$m), which agree, within the measurement uncertainties, with the values predicted for DdDm-1's temperature and density by the models presented in Table~\ref{fluxes} and discussed in {\S\ }3.

Orders were trimmed and merged and line fluxes for DdDm-1 were measured with SMART\footnote{SMART was developed by the IRS Team at Cornell University and is available through the Spitzer Science Center at Caltech.} \citep{Higdon04}, which produces fluxes and uncertainty estimates for each line from its line-fitting routine. SMART was also able to fit a thermal continuum to the SH-LH spectrum, obtaining a temperature of 125~K, typical of thermal dust emission. Figs.~\ref{sl1} and \ref{lh} show the IRS spectra.

\subsection{Echelle Data and the Radial Velocity of DdDm-1}

DdDm-1 was observed in June 2002 using the echelle spectrograph on the Mayall 4-meter telescope at Kitt Peak. We used the T2KB 2048x2048 pixel CCD, binned 2x2. We observed in two configurations in order to cover the full spectral range: the blue configuration, which spans wavelengths between approximately 4300 \AA~ and 7200 \AA, and the red configuration, with a spectral range between 6500 \AA~ and 9600 \AA. Total exposure time was 6600 s. We used the 79-63$^\circ$ echelle grating and the 226-2 cross disperser on all nights. The reductions were done using the {\it echelle} package in IRAF. Though accurate flux calibration among the observed echelle orders proved impossible, we were able to extract kinematic information from the observations. 

Observed and measured wavelengths for a selection of lines in the spectrum are given 
in Table~\ref{radvel}; based on these measurements we find the radial velocity of 
DdDm-1 to be -300.9 $\pm$ 1.4 km s$^{-1}$. We used the {\it rvcorrect} task in the {\it astutil} package in IRAF to correct for the earth's rotation and orbital motion at the time of the observations and found a correction of -9.4 km s$^{-1}$, giving a heliocentric radial velocity of -310.3 $\pm$ 1.4 km s$^{-1}$. The radial velocity of DdDm-1 has been measured by \citet{BC84} to be -304$\pm$ 20 km s$^{-1}$; and by \citet{WCP05} as -317 $\pm$ 13 km s$^{-1}$; our value agrees very well with both of these, and is better constrained.

We take the opportunity here to mention that the echelle spectrum contains many forbidden iron lines that appear split. We will address the issue of DdDm-1's expansion and ramifications for its morphology in a future paper.

\section{ANALYSIS}

We performed both an empirical and a numerical analysis of DdDm-1, using the line strengths reported in the previous section and listed in Table~\ref{fluxes}. We first employed the abundance software package  ELSA \citep{J07}, a new C program based upon a 5-level atom routine, to derive empirical electron temperatures and densities as well as ionic and elemental abundances of numerous elements. These same abundances were then used as input for detailed photoionization model calculations of DdDm-1 using the program Cloudy \citep{Fer98} version~07.02.00. The purpose of this numerical work was to further refine the empirical abundances to produce our final abundance set as well as to derive information about central star properties. These two steps are discussed separately below. This dual-phase approach generated a final set of elemental abundances for DdDm-1 which is very reliable and allows us to compare results from numerical and empirical methods. We now describe each of the steps in detail.

\subsection{Empirical Analysis\label{ea}}

In this step, strengths of many of the emission lines in Table~\ref{fluxes} were entered as input in ELSA. The program then derived electron temperature and density estimates based upon temperature-sensitive or density-sensitive line sets and then calculated ionic abundances. Finally, total elemental abundances were determined through the use of ionization correction factors which account for the contributions of unobserved ions to the total. Results for electron temperatures and densities, ionic abundances, and elemental abundances derived using the empirical method are presented in Tables~\ref{td}, \ref{ions}, and \ref{elements}, respectively.

In Table~\ref{td} we report the values of five electron temperatures from diagnostic emission line ratios of [O~III], [N~II], [O~II], [S~II], and [S~III], and three values of electron density, [S~II], [Cl~III], and [S~III]. (The emission lines used to calculate these values are provided in a footnote to the table.) Note that for the [S~III] temperature the $\lambda$9532 line strength was used, since the line strength ratio of $\lambda$9532/$\lambda$9069 line exceeded the theoretical value, thereby indicating that $\lambda$9069 emission has been partially absorbed in the Earth's atmosphere. 

Columns~3-6 list temperature and density values derived by \citet{C87}, \citet{BC84}, \citet{HKB04}, and \citet{WLB05}. All of the [O~III] temperatures agree closely with our new value. However, there is nearly a 2000~K range in the [N~II] temperatures with larger variations still for the [O~II], [S~II], and [S~III] temperatures, although note that our values for the first two are very consistent with those computed by \citet{WLB05}.  At the same time all three of the electron densities which we derived agree nicely with one another as well as with the values inferred by earlier studies.

Empirical ionic abundances based upon our temperature and density values in Table~\ref{td} are shown in Table~\ref{ions}. For each ion indicated in column~1 we list the electron temperature in column~2 that was used to determine the ion abundance given in column~3. At the same time, the [S~II] density was used for all ionic abundance calculations. Uncertainties were determined by adding in quadrature the individual contributions to uncertainty made by such things as temperature and density uncertainties as well as uncertainties in the line strengths themselves. For most ions we provide several abundance values, where each is based upon the emission line whose wavelength is indicated in parentheses in column~1. When more than one abundance is computed for an ion, the last value is a weighted mean of the values marked with an asterisk (*), where the weight is related to the uncertainty assigned to each of the individual values for that ion. The last entry for each element is the value of the ionization correction factor that was used to compute the total elemental abundance. The ICFs were determined using the relations provided in \citet{KH01}.

For comparison purposes we have included results from \citet{C87} and \citet{BC84} for ions and emission lines provided in those two studies. [An additional comparison in the cases of Ne and S will be made with the results of \citet{D03} below.] For the major ions there appear to be no significant discrepancies among the three studies. The situation is nearly the same for the ICFs, although the moderate variance among derived values for N suggests that the abundance of this major element may be in dispute.

\subsection{Numerical Analysis\label{na}}

We next employed the program Cloudy \citep{Fer98}, version 07.02.01 to calculate detailed photoionization models of DdDm-1 with the goal of refining our empirically derived abundances as well as inferring information about the central star properties. Cloudy uses a trial set of input parameters, whose values are determined by the investigator, and predicts emission line strengths and physical conditions for the nebula. In refining the empirical abundances, then, we followed the procedure described and used by \citet{KH98} to study DdDm-1 previously, a routine which has proven to produce a set of robust results. The steps in the procedure are as follows:

\begin{enumerate}

\item Calculate a photoionization model whose output line strengths closely match the observed ones of the real nebula.

\item Derive a set of  empirical abundances for the model nebula using the output line strengths and ELSA.

\item Use the model empirical abundances from 2., along with the model input abundances, to determine a correction factor for each element, where the correction factor is the ratio of the model input abundance to the model empirical abundance.

\item Multiply the empirical abundance values determined in {\S}\ref{ea} by the relevant correction factors to obtain a final set of abundances.

\end{enumerate}

In calculating the model in step~1, our empirically derived abundances from Table~\ref{elements} and electron density from Table~\ref{td} were used to set the input parameter values in the first model. The predicted emission line strengths of that model were then compared with their observed counterparts, the parameter values adjusted accordingly, and a new model calculated. This process was repeated until a suitable match between theory and observation was obtained. Note that the images of DdDm-1 presented above support our use of a relatively simple density distribution for our modeling exercise, since they suggest a smooth distribution of matter and a symmetrical shape. We present the results for two successful models, 18 and 32, in Table~\ref{fluxes}. 

For model~18 the input central star flux was taken from the H-Ni grid of synthetic central star fluxes by \citet{R02}, which was calculated by assuming non-LTE hydrostatic conditions, line-blanketing, and plane-parallel geometry. The luminosity of the central star was taken to be 1000~L$_{\odot}$. The model was radiation bounded with a constant total density of 4000~cm$^{-3}$ throughout the nebula, consistent with the smooth appearance of the nebula in Figs.~\ref{image1} and \ref{image2} and the electron density reported in Table~\ref{td}; the filling factor was unity. Model~18 was successful in reproducing most of the important lines (shown in bold) in Table~\ref{fluxes}. However, the predicted strength of [O~III] $\lambda$4363 was somewhat higher than the observed value, while the [S~II] $\lambda\lambda$6716,6731 line strengths were also overpredicted by the model. Numerous models were run in an effort to reduce these particular problems, but improvements in these lines came only with serious damage to other predicted line strengths; in the case of sulfur this often meant poor matches with the [S~III] and [S~IV] lines. An example is our attempt to reduce [S~II] emission by truncating the nebula, i.e. making it matter bounded with the intention of reducing the volume of gas in the outer region of the nebula where large amounts of [S~II] are produced. However, this led to a significant reduction of emission from ions such as [O~II] and [N~II], making the altered model an untenable solution.

Model 32 differs from 18 most significantly in the character of the central star. For this model we used a blackbody spectrum of T$_{eff}$=40,000K with a bolometric luminosity of 10$^5$~L$_{\odot}$. The other major difference was that model~32 had a filling factor of 0.5. Otherwise the density was unchanged and the abundances were very similar to those employed in model~18. For this model, the prediction of [O~III] $\lambda$4363 is slightly better, although this time it is below rather than above the observed value. In addition, the predicted strength of [O~III] $\lambda$5007 is below the observed level. Model~32 was primarily an attempt to achieve improved agreement in the [S~II] lines over that found in model~18. However, in doing so the level of agreement in the near IR lines of [S~III] and [S~IV] at 10.5$\mu$m became worse. 

The main parameters for models 18 and 32 are summarized in Table \ref{parameters}. For the remainder of the analysis we will use model~18, since it includes the use of a realistic central star model spectrum, and its predicted line strengths satisfactorily match most of the important observed line strengths. In addition, this model's H$\beta$ luminosity closely matches the observed value. The fact that this model does not reproduce a few line strengths exactly is not a problem here, since we are using the model results primarily to derive a correction factor (see points~3 and 4 above).

\section{DISCUSSION}

\subsection{Adopted Abundances of DdDm-1}

The correction factors (based upon model~18) and our final abundance results for DdDm-1 are presented in column~3 of Table~\ref{final}. Final  elemental abundances relative to H were determined by adding the log of the correction factor in column~2 to the associated value in Table~\ref{elements}. The uncertainties given in column~3 are statistical and based upon error propagation results calculated by ELSA from estimated line strength errors. Columns 4-8 of Table~\ref{final} show values for comparison purposes from \citet{C87}, \citet{BC84}, and \citet{WLB05} for DdDm-1, \citet{AGS05} for the sun, and \citet{E98} for the Orion Nebula, respectively. The last column provides a comparison of our abundances in column~3 with solar abundances, using the usual bracket notation defined in the footnote. Note that since neither Si nor Fe is currently included in ELSA, our final values for these elements correspond to the model input values required to reproduce several of the measured line strengths of these elements.

The fact that all of the correction factor values in column~2 exceed unity suggests that the empirical method of abundance determinations tends to underestimate the abundance of each element. This is possibly caused by insufficient corrections for unobserved ionization stages by the ionization correction factors used in the empirical method. Furthermore, the magnitude of the offsets is on the order of 0.10-0.15~dex, or roughly the size of the uncertainties which we established for the final abundances.

There is good agreement among the four studies of DdDm-1 represented in Table~\ref{final} for He/H, O/H, Ne/H, Si/H, and S/H, where all ratios but the first one represent alpha elements, and therefore their abundances are expected to exhibit lockstep behavior. We note that our value for O/H tends to be slightly higher than the others, although even this difference is likely explained by uncertainties. At the same time our value for Fe/H agrees reasonably well with that published by \citet{C87}. 

While our N/H abundance is within a factor of 2 of those found in the other two studies, C/H is markedly discrepant among the four determinations, where the range exceeds two orders of magnitude. Our value is more than a factor of two lower than the recently measured level published by \citet{WLB05}. We note that \citet{BC84} determined their C abundance using the recombination line C~II $\lambda$4267, and it is often the case that abundances derived from recombination lines yield significantly higher values than abundances determined from collisionally excited lines (\citep{WLB05}. Unlike Barker \& Cudworth, we did not detect 4267 in our spectrum, nor apparently did \citet{WLB05}. However, an upper limit for the $\lambda$4267 line strength in our spectrum is 0.03 (H$\beta$=100) or about 1/10 the strength reported by \citet{BC84}. The correction for reddening is insignificant. Running this value through ELSA produces an upper limit on the abundance ratio of C$^{+2}$/H$^+$ of 3.1E-5, or about 1/10 the level of this ratio determined by \citet{BC84}. In any case, our C/H was inferred directly from our photoionization models, using the 1909 line as a constraint, and thus we consider it somewhat uncertain. In terms of C/O, our result and that of \citet{C87} and \citet{WLB05} suggest that DdDm-1 is a C-poor (or O-rich) system (C/O$<$1), while \citet{BC84}'s value implies a C-rich (O-poor) system (C/O$>$1).  

\citet{D03} used infrared and optical line measurements acquired at the IRTF and the 2.7m telescope at McDonald Observatory, respectively, to study abundances of S and Ne in DdDm-1. Table~\ref{dinerstein} provides a comparison of their results with ours. The agreement between the two groups is remarkably good, with the exception of the factor of 4 discrepancy in the case of Ne$^+$/H$^+$. We are currently unable to explain this disagreement, as our measured strength of [Ne II]~12.8$\mu$m
agrees closely with the value reported by \citet{D03}, as do our derived electron temperature and density values. We also employed a collision strength of 0.318 \citep{GMB01} which agrees closely with their value of 0.306 taken from \citet{JK87}.  On the other hand, the close agreement in the case of sulfur is strong evidence that the ionization stages above S$^{+3}$ in DdDm-1 are relatively unpopulated.

Fig.~\ref{abun_sun} presents a comparison of the elemental abundances of DdDm-1 and three other well-studied halo PNe, BB1, H4-1, and K648 [each was analyzed by \citet{HKB04}, from which the abundances in the figure are taken], all normalized to solar values from \citet{AGS05}. For clarity, uncertainties are not plotted; they generally have values of 0.10~dex or less. 

We can see very clearly that all four PNe are metal-poor, since the offsets for all of the alpha elements (O through Ar) are negative by significant amounts. However, it is interesting that for any one object these offsets do not have the same value for the all of the alpha elements as would be expected from nucleosynthesis theory\footnote{Since alpha elements result from He burning processes, one expects them to track each other, and therefore the offsets should be roughly the same for any one object.}. For example we see in column~8 of Table~\ref{final} that for DdDm-1 these values vary from -0.25 for Ar/H to to -1.36 for Si/H.  Some of the variation can certainly be explained by uncertainty, particularly in the cases of Ar and Cl where line strengths are weak and only one or two ionization states have observable lines. It is likely that some of the underabundance of Si in particular is the result of dust depletion, as this element is highly refractory \citep{SS96}. In fact its offset from solar is essentially identical to that of Fe, another refractory element. However, dust cannot explain the large offset differences between O and Ne for H4-1 and BB1, each long known for their unusual Ne abundances relative to O. In addition there are large differences in S offsets for BB1, H4-1, and K648, perhaps related to the S anomaly discussed below. Finally, the Ar offset is well above its expected value for K648 but well below the expected one for BB1. It is unlikely that any of these peculiar offsets can be explained by dust formation. Rather they may be related to the increased scatter often seen in low metallicity halo stars [see the data compilation in Fig.~1.2 of \citet{M03}]. It is hoped that with future discoveries of additional halo PNe this situation will become better understood.

We also see in Fig.~\ref{abun_sun} that C/H and C/O are much lower in DdDm-1 than in the other three halo PNe, with significantly subsolar values for both ratios. In contrast, the C/O offset is positive for the other three objects, with a very high value in the case of BB1. 

As alpha elements, S and O are expected to evolve in lockstep, as we explained above.
\citet{HKB04} found that while this is true when H~II regions and blue compact galaxies are used to probe the abundances of these two elements, the expectation is often unmet in the case of PNe\footnote{See also the discussion of Ne and S abundances in H~II regions in \citet{LBS07}.} These authors found marked scatter in S abundances for  PNe of roughly the same O abundance. In addition, S abundances were regularly determined to be below the expected level for a given O abundance by 0.3~dex on average. This unexpected finding was referred to as the sulfur anomaly by \citet{HKB04}. In the specific case of DdDm-1, however, \citet{HKB04} found that its S abundance was close to the value expected from its O abundance.

We now revisit this situation with our updated S and O abundances obtained here.
Employing our O abundance in Table~\ref{elements} along with a least squares fit to measurements of 12+log(S/H) versus 12+log(O/H) in H~II regions in M101 \citep{KBG03} and blue compact galaxies \citep{IT99} to estimate the expected S abundance\footnote{12+log(S/H)=0.888[12+log(O/H)] - 0.683}, we obtain 12+log(S/H)=6.60 as the expected value for S. Our corresponding measured value is 6.47, yielding a sulfur deficit of 6.60-6.47=0.13, which is close to the uncertainty in the S abundance and less than the typical value of 0.3 for the sulfur deficit found by \citet{HKB04}.  We conclude that the S and O abundances associated with DdDm-1 are consistent with the expected lockstep behavior for these two elements.

It has been suggested by \citet{PBS06} that the sulfur anomaly found by \citet{HKB04} could be the result of dust formation. In particular, S may be removed by the formation of compounds such as MgS and FeS in those PNe exhibiting large S deficits. In fact this would be expected to occur more readily in C-rich environments, where sulfide formation is favored. Interestingly, the low C abundance which we find for DdDm-1, with C/O$<$1, implies that oxygen-rich chemistry exists in the nebula of DdDm-1 and that dust composition should be dominated by silicates and other oxygen-rich species and not sulfides. Thus, if the sulfur anomaly is indeed related to sulfide formation, then the small S deficit that we observe in DdDm-1 would be expected as the result of its low value for C/O.

\begin{figure}
   \includegraphics[width=4in,angle=270]{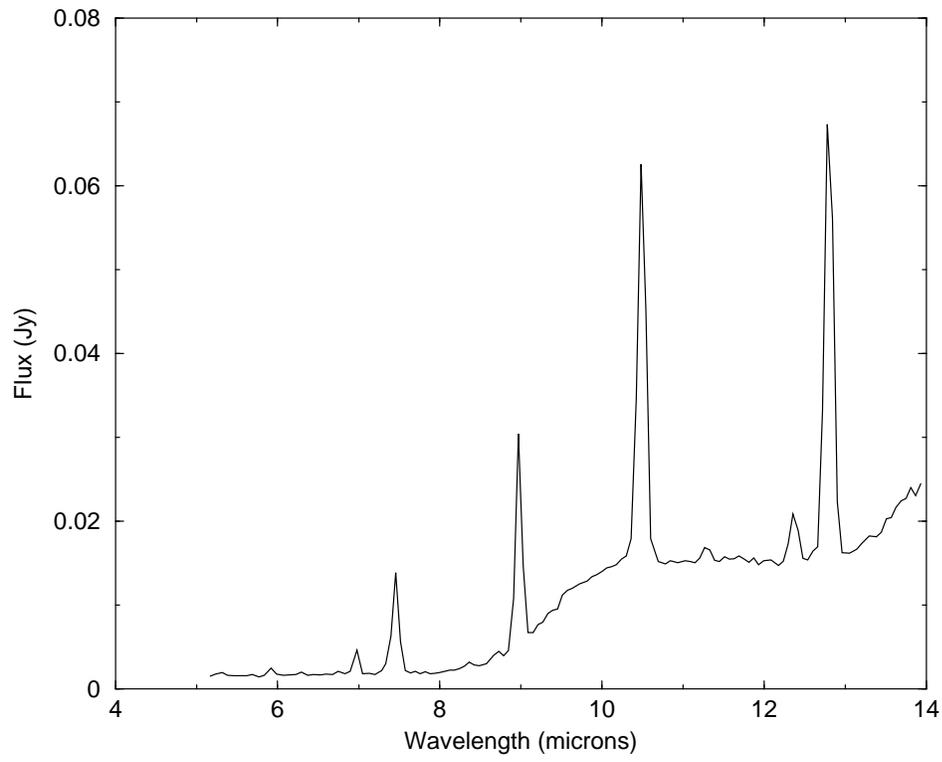} 
   \caption{IRS merged SL2-SL1 spectrum}
   \label{sl1}
\end{figure}

\begin{figure}
   \includegraphics[width=4in,angle=270]{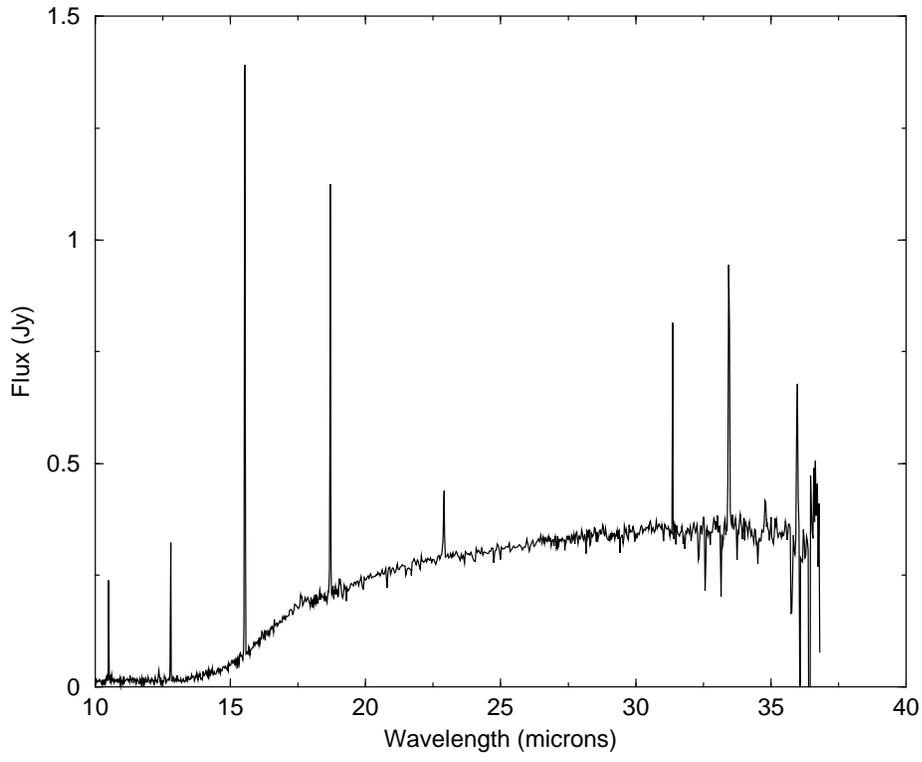} 
   \caption{IRS merged SH-LH spectrum}
   \label{lh}
\end{figure}

\begin{figure}
   \includegraphics[width=4in,angle=270]{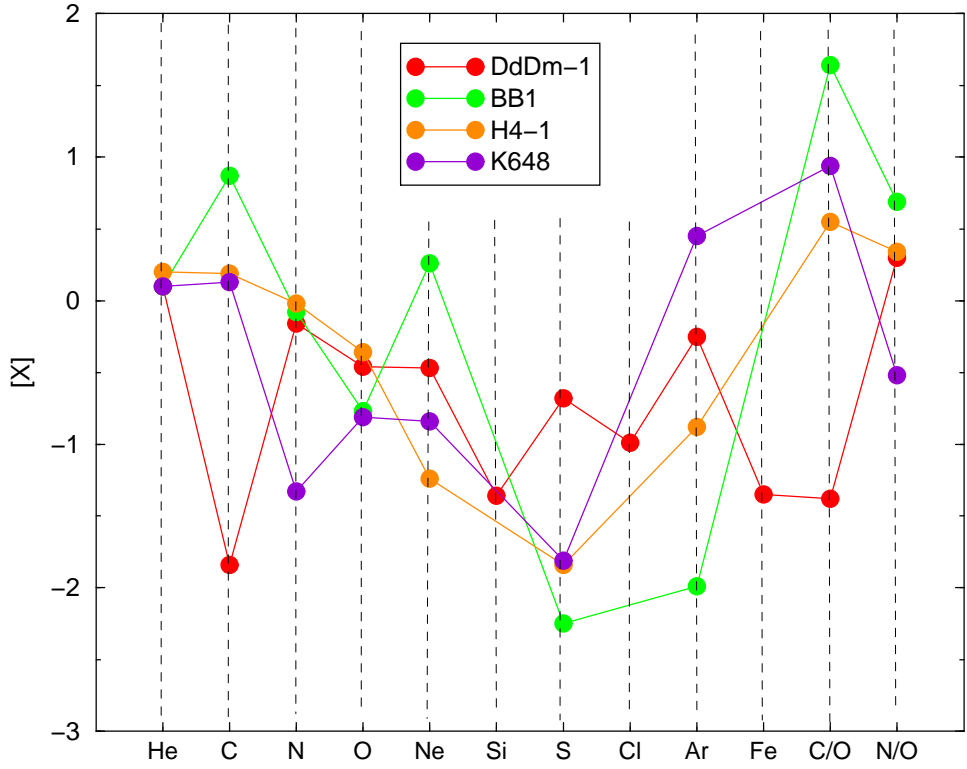} 
   \caption{Plot of [X] versus element ratio for DdDm-1 (red), BB1 (green), H4-1 (orange), K648 (violet), where [X] is the logarithmic value normalized to solar of a ratio on the horizontal axis.}
   \label{abun_sun}
\end{figure}

Further evidence for the low C/O ratio in DdDm-1 is the absence of polycyclic aromatic hydrocarbon (PAH) emission bands in the IRS SL spectra (Fig.~\ref{sl1}).Ê The IDL routine PAHFIT \citep{SD07} was applied to the SL spectra, and no evidence was found of the PAH emission features near 3.3, 6.2, 7.7, 8.7, and 11.3 microns, which are strong in the IR spectra of many PNe \citep{CB05}.Ê In their study of ISO spectra of 43 PNe, Cohen and Barlow found that 17 objects exhibited strong PAH emission and that the 7.7 and 11.3 micron PAH band strengths relative to the total infrared luminosity are correlated with the nebular C/O ratio.Ê As is evident from our IRS spectra in Figures \ref{sl1} and \ref{lh}, all of the emission features seen in DdDm-1 are nebular lines with no broad PAH emission evident.Ê However, a strong IR continuum, apparently due to warm dust emission, is evident.Ê Blackbody fits to the LH spectrum using SMART give a good fit for a temperature of 125(+/- 7) K.Ê This is comparable to other PNe studied by ISO and Spitzer.Ê Finally, no significant silicate absorption bands at 9.7 and 18 microns are visible in the IR spectra.Ê These findings further support our result for a low C/O ratio in DdDm-1, as well as the liklihood that Si/O is low as well.

\subsection{Central Star Properties}

The central star temperature which we infer for DdDm-1 is 55,000~K with a luminosity of 10$^3$~L$_{\odot}$, based upon our preferred model~18 (see Table~\ref{parameters}). Recall that model~18 employed a stellar spectrum from \citet{R02} which was calculated by assuming realistic conditions of high gravity and low metallicity for the central star. Our effective temperature is somewhat higher than the value of 40,000~($\pm$5,000)~K determined by \citet{PTPR92} and 45,600~K estimated from models by \citet{HHM97}. This situation may reflect our use of model stellar fluxes calculated specifically for a low metallicity regime.

These derived central star properties along with the O abundance for DdDm-1 can be compared with AGB star model tracks calculated by \citet{VW94} in order to infer a progenitor mass. Data in their Fig.~7 suggests that the central star of DdDm-1 is a He-burning object which had a main sequence mass of $<$1~M$_{\odot}$. This relatively low mass is consistent with the idea that DdDm-1 is associated with an old stellar population such as that found in the halo.

\section{SUMMARY \& CONCLUSIONS}

We have reported on new IR and UV spectra of the halo planetary nebula DdDm-1 obtained with the Spitzer Space Telescope with the IRS and Hubble Space Telescope with the FOS, respectively. By combining these new data with existing optical measurements, the nebular abundances of He, C, N, O, Ne, Si, S, Cl, Ar, and Fe were determined. The abundance analysis included the computation of detailed photoionization models which made use of an input stellar atmosphere whose characteristics were consistent with a central star of low metallicity. We also present new echelle data for DdDm-1. We have found the following:

\begin{enumerate}

\item We present what we believe to be the first resolved image of DdDm-1. Reconstructed imagery taken with the HST in 1993 indicated that the morphology of DdDm-1
can be described as two orthogonal elliptical components in the central part surrounded by an extended halo.
The extent of the emission is somewhat larger than was previously reported in the literature.

\item We present new UV, IR, and echelle data, where the first two sets were acquired using HST and Spitzer, respectively.

\item Our new determinations of the Si and Fe abundances indicate that their levels are far below those expected from the metallicity of DdDm-1, possibly indicating that their gas phase abundances have been depleted by dust formation.

\item We determine that C/O $<$ 1, although our C/H abundance is uncertain. However, a C/O ratio below unity suggests that the chemistry of the nebula is O-rich in character. Thus, sulfides should be absent any dust that has formed in the environment of DdDm-1.

\item We find that the abundance of S$^{+3}$ which we determined directly from our new [S~IV] 10.5$\mu$m measurement agrees closely with another modern one by \citet{D03} and is consistent with the level predicted by the value of the ICF obtained when only optical lines of [S~II] and [S~III] are used. The small abundance of S$^{+3}$ that we derive indicates that DdDm-1 is a relatively low excitation nebula when compared with other PNe.

\item Our total gas-phase S abundance for DdDm-1 is consistent with the value expected from its O abundance, under the assumption of lockstep behavior between these two elements. Thus DdDm-1 has a negligible S deficit. On the other hand, if the large S deficits observed in other PNe are indeed due to sulfide formation in a C-rich environment as others have suggested, then the small S deficit of DdDm-1 is entirely consistent with its C-poor properties.

\item For the central star we find that T$_{eff}$=55,000~K and L=1000~L$_{\odot}$. Comparing the star with theoretical model tracks suggests that it is a He-burning object with a mass of less than 1~M$_{\odot}$.

\item The heliocentric radial velocity of DdDm-1 is -310.3 $\pm$ 1.4 km s$^{-1}$.

\end{enumerate}

\acknowledgments

All four authors thank their respective institutions for travel support for attending IAU Symposium 234 on planetary nebulae in spring, 2006, where we had the opportunity to meet and discuss various problems surrounding this and related studies of target objects. RJD and KBK wish to thank Rice graduate student Greg Brunner for applying PAHFIT and SMART blackbody analyses to our Spitzer IRS spectra.
RBCH acknowledges fruitful discussions with Angela Speck about dust formation in PNe. KBK thanks R. Gruendl and Y.-H. Chu for considerable advice and assistance with the echelle observations. RBCH is also grateful for partial support of his research by NSF grant AST 03-07118 to the University of Oklahoma.

\begin{deluxetable}{lcr@{}lr@{}lcc}
\tabletypesize{\normalsize}
\setlength{\tabcolsep}{0.07in}
\tablecolumns{8}
\tablewidth{0in}
\tablecaption{Fluxes and Intensities\label{fluxes}}
\tablehead{
\colhead{Line} &
\colhead{f($\lambda$)} &
\multicolumn{2}{c}{F($\lambda$)} &
\multicolumn{2}{c}{I($\lambda$)} &
\colhead{Model 18} &
\colhead{Model 32}}
\startdata
He II $\lambda$1640 & 1.136 & 2.21 &:  & 2.58 & $\pm$1.56:  & 0.040 & 0.123\\
O III] $\lambda$1662 & 1.129 & 5.76 & & 6.72 & $\pm$3.58 & 5.18 & 2.01\\
N III] $\lambda$1750 & 1.119 & 1.10 &::  & 1.28 & $\pm$0.93::  & 18.4 & 8.0\\
Si III] $\lambda$1887 & 1.200 & 4.95 & & 5.84 & $\pm$3.27 & 16.1 & 9.85\\
{\bf C III] $\lambda$1909} & 1.229 & 11.0 & & 13.0 & $\pm$7.44 & 8.21 & 7.60\\
\[[O II] $\lambda$2470 & 1.025 & 9.58 & & 11.0 & $\pm$5.43 & 12.1 & 10.9\\
{\bf\[[O II] $\lambda$3727} & 0.292 & 103 & & 107 & $\pm$24 & 116 & 111\\
He II + H11 $\lambda$3770 & 0.280 & 3.57 & & 3.71 & $\pm$0.81 & 3.90 & 4.04\\
He II + H10 $\lambda$3797 & 0.272 & 4.50 & & 4.67 & $\pm$1.01 & 5.22 & 5.39\\
He II + H9 $\lambda$3835 & 0.262 & 6.76 & & 7.01 & $\pm$1.49 & 7.23 & 7.44\\
{\bf \[[Ne III] $\lambda$3869} & 0.252 & 29.4 & & 30.4 & $\pm$6.40 & 32.3 & 21.3\\
He I + H8 $\lambda$3889 & 0.247 & 19.0 & & 19.6 & $\pm$4.09 & 18.7 & 12.7\\
\[[Ne III] $\lambda$3968 & 0.225 & 11.3 &\tablenotemark{a} & 11.7 & $\pm$5.58\tablenotemark{a} & 9.73 & 6.42\\
H$\epsilon$ $\lambda$3970 & 0.224 & 15.6 &\tablenotemark{a} & 16.0 &\tablenotemark{a} & 15.81 & 16.2\\
He I + \[[Fe III] $\lambda$4008 & 0.214 & 0.591 & & 0.609 & $\pm$0.120 & \nodata & \nodata\\
He I + He II $\lambda$4026 & 0.209 & 2.12 & & 2.18 & $\pm$0.43 & 2.23 & 2.12\\
\[[Fe III] $\lambda$4046 & 0.203 & 0.102 &::  & 0.105 & $\pm$0.055:: & \nodata & \nodata\\
\[[S II] $\lambda$4071 & 0.196 & 2.67 & & 2.74 & $\pm$0.53 & 3.79 & 1.62\\
H$\delta$ $\lambda$4101 & 0.188 & 25.4 & & 26.1 & $\pm$4.95 & 25.8 & 26.3\\
He I $\lambda$4121 & 0.183 & 0.227 &::  & 0.233 & $\pm$0.122:: & 0.273 & 0.242\\
He I $\lambda$4144 & 0.177 & 0.280 &::  & 0.287 & $\pm$0.150:: & 0.321 & 0.303\\
C III $\lambda$4167 & 0.170 & 0.104 & & 0.106 & $\pm$0.020 & \nodata & \nodata\\
H$\gamma$ $\lambda$4340 & 0.124 & 47.7 & & 48.5 & $\pm$8.32 & 47.0 & 47.2\\
{\bf \[[O III] $\lambda$4363} & 0.118 & 5.07 & & 5.15 & $\pm$0.88 & 8.03 & 3.87\\
He I $\lambda$4388 & 0.112 & 0.607 & & 0.616 & $\pm$0.104 & 0.581 & 0.552\\
He I $\lambda$4472 & 0.090 & 4.91 & & 4.97 & $\pm$0.81& 4.81 & 4.57\\
\[[Fe II-III] $\lambda$4606 & 0.056 & 0.139 &::  & 0.140 & $\pm$0.072:: & 0.126 & 0.121\\
\[[Fe III] $\lambda$4658 & 0.043 & 2.38 & & 2.39 & $\pm$0.36 & 2.14 & 2.05\\
\[[Fe III] $\lambda$4702 & 0.032 & 0.793 & & 0.796 & $\pm$0.118 & 0.733 & 0.705\\
He I + \[[Ar IV] $\lambda$4711 & 0.030 & 0.793 & & 0.796 & $\pm$0.118 & 1.03 & 0.981\\
\[[Fe III] $\lambda$4734 & 0.024 & 0.312 & & 0.313 & $\pm$0.046 & 0.247 & 0.239\\
\[[Ar IV] $\lambda$4740 & 0.023 & 0.166 & & 0.167 & $\pm$0.024 & 0.193 & 0.134\\
\[[Fe III] $\lambda$4755 & 0.019 & 0.398 & & 0.400 & $\pm$0.058 & 0.391 & 0.375\\
\[[Fe III] $\lambda$4770 & 0.015 & 0.280 & & 0.281 & $\pm$0.041 & 0.246 & 0.236\\
\[[Fe III] $\lambda$4778 & 0.013 & 0.149 & & 0.149 & $\pm$0.021 & 0.117 & 0.114\\
H$\beta$ $\lambda$4861 & 0.000 & 100 & & 100 & & 100 & 100\\
\[[Fe III] $\lambda$4881 & -0.012 & 0.989 & & 0.987 & $\pm$0.137 & 0.694 & 0.683\\
He I $\lambda$4922 & -0.021 & 1.20 & & 1.20 & $\pm$0.16 & 1.23 & 1.16\\
\[[O III] $\lambda$4959 & -0.030 & 150 & & 149 & $\pm$20 & 163 & 108\\
{\bf \[[O III] $\lambda$5007} & -0.042 & 458 & & 455 & $\pm$61 & 490 & 326\\
Si II $\lambda$5056 & -0.053 & 5.20(-2) &::  & 5.16 & $\pm$2.62(-2):: & \nodata & \nodata\\
\[[Fe III] $\lambda$5085 & -0.060 & 0.436 &:  & 0.432 & $\pm$0.135: & 0.050 & 0.049\\
\[[Fe II] $\lambda$5159 & -0.077 & 0.717 & & 0.709 & $\pm$0.092 & \nodata & \nodata\\
\[[Ar III] $\lambda$5192 & -0.085 & 0.914 & & 0.903 & $\pm$0.116 & 0.167 & 0.104\\
\[[N I] $\lambda$5199 & -0.086 & 0.388 & & 0.383 & $\pm$0.049 & 0.427 & 0.048\\
\[[Fe III] $\lambda$5270 & -0.102 & 1.18 & & 1.16 & $\pm$0.15 & 1.14 & 1.11\\
\[[Cl III] $\lambda$5518 & -0.157 & 0.217 &:  & 0.212 & $\pm$0.066: & 0.161 & 0.128\\
\[[Cl III] $\lambda$5538 & -0.161 & 0.271 &:  & 0.265 & $\pm$0.082: & 0.190 & 0.152\\
\[[N II] $\lambda$5755 & -0.207 & 1.43 & & 1.39 & $\pm$0.17 & 1.67 & 1.44\\
{\bf He I $\lambda$5876} & -0.231 & 14.6 & & 14.1 & $\pm$1.76 & 14.2 & 12.9\\
O I $\lambda$6046 & -0.265 & 0.103 &::  & 9.93 & $\pm$5.03(-2):: & \nodata & \nodata\\
\[[O I] $\lambda$6300 & -0.313 & 2.47 & & 2.37 & $\pm$0.32 & 2.95 & 0.425\\
\[[S III] + He II $\lambda$6312 & -0.315 & 1.84 & & 1.76 & $\pm$0.24 & 2.55 & 1.21\\
Si II $\lambda$6347 & -0.322 & 8.45(-2) & & 8.09 & $\pm$1.09(-2) & \nodata & \nodata\\
\[[O I] $\lambda$6364 & -0.325 & 0.861 & & 0.824 & $\pm$0.112 & 0.940 & 0.136\\
\[[N II] $\lambda$6548 & -0.358 & 17.0 & & 16.2 & $\pm$2.29 & 20.7 & 18.8\\
H$\alpha$ $\lambda$6563 & -0.360 & 296 & & 282 & $\pm$1 & 289 & 287\\
{\bf \[[N II] $\lambda$6584} & -0.364 & 54.0 & & 51.4 & $\pm$7.34 & 61.0 & 55.4\\
He I $\lambda$6678 & -0.380 & 3.84 & & 3.65 & $\pm$0.53 & 3.30 & 3.08\\
{\bf \[[S II] $\lambda$6716} & -0.387 & 3.02 & & 2.86 & $\pm$0.42 & 6.73 & 2.00\\
{\bf \[[S II] $\lambda$6731} & -0.389 & 5.13 & & 4.86 & $\pm$0.72 & 10.8 & 3.38\\
\[[Ar V] $\lambda$7006 & -0.433 & 9.33(-2) & & 8.79 & $\pm$1.40(-2) & \nodata & \nodata\\
He I $\lambda$7065 & -0.443 & 8.01 & & 7.54 & $\pm$1.22 & 8.98 & 10.8\\
\[[Ar III] $\lambda$7136 & -0.453 & 8.10 & & 7.61 & $\pm$1.25 & 10.5 & 8.15\\
\[[Fe II] $\lambda$7155 & -0.456 & 9.95(-2) &:  & 9.35 & $\pm$3.06(-2):  & \nodata & \nodata\\
\[[Ar IV] + \[[Fe II] $\lambda$7170 & -0.458 & 3.59(-2) &:  & 3.37 & $\pm$1.10(-2): & 0.006 & 0.004\\
O I $\lambda$7255 & -0.471 & 9.52(-2) & & 8.92 & $\pm$1.51(-2) & \nodata & \nodata\\
He I $\lambda$7281 & -0.475 & 0.927 & & 0.869 & $\pm$0.148 & 1.02 & 0.854\\
{\bf \[[O II] $\lambda$7324} & -0.481 & 16.4 & & 15.3 & $\pm$2.63 & 14.4 & 13.0\\
\[[Ni II] $\lambda$7378 & -0.489 & 4.85(-2) &:  & 4.54 & $\pm$1.51(-2): & \nodata & \nodata\\
\[[Fe II] $\lambda$7388 & -0.490 & 2.25(-2) & & 2.10 & $\pm$0.37(-2) & \nodata & \nodata\\
C III $\lambda$7578 & -0.516 & 5.30(-3) & & 4.94 & $\pm$0.90(-3) & \nodata & \nodata\\
\[[Cl IV] $\lambda$7531 & -0.510 & 1.73(-2) &:   & 1.62 & $\pm$0.54(-2):  & 0.101 & 0.084\\
\[[Ar III] $\lambda$7751 & -0.539 & 1.95 & & 1.81 & $\pm$0.34 & 2.53 & 1.97\\
\[[Ni III] $\lambda$7890 & -0.556 & 8.25(-2) &:  & 7.65 & $\pm$2.62(-2): & \nodata & \nodata\\
P16 $\lambda$8467 & -0.618 & 0.394 & & 0.362 & $\pm$0.077 & 0.496 & 0.540\\
\[[Cl III] $\lambda$8481 & -0.620 & 0.134 & & 0.123 & $\pm$0.026 & 0.008 & 0.006\\
\[[Cl III] + P15 $\lambda$8501 & -0.622 & 0.525 & & 0.482 & $\pm$0.103 & 0.584 & 0.630\\
P14 $\lambda$8545 & -0.626 & 0.620 & & 0.569 & $\pm$0.123 & 0.512 & 0.537\\
\[[Cl II] $\lambda$8579 & -0.629 & 0.142 & & 0.130 & $\pm$0.028 & 0.112 & 0.076\\
P13 $\lambda$8598 & -0.631 & 0.706 & & 0.648 & $\pm$0.141 & 0.634 & 0.665\\
\[[Fe II] $\lambda$8617 & -0.633 & 0.131 & & 0.120 & $\pm$0.026 & 0.288 & 0.077\\
P12 $\lambda$8665 & -0.637 & 0.899 &::  & 0.824 & $\pm$0.442:: & 0.794 & 0.831\\
P11 $\lambda$8750 & -0.644 & 1.14 & & 1.04 & $\pm$0.23 & 1.01 & 1.06\\
P10 $\lambda$8863 & -0.654 & 1.41 & & 1.29 & $\pm$0.29 & 1.32 & 1.37\\
P9 $\lambda$9015 & -0.666 & 1.99 & & 1.82 & $\pm$0.42 & 1.76 & 1.83\\
{\bf \[[S III] $\lambda$9069} & -0.670 & 20.7 & & 18.9 & $\pm$4.35 & 20.3 & 11.8\\
P8 $\lambda$9228 & -0.610 & 3.40 & & 3.13 & $\pm$0.66 & 2.43 & 2.52\\
{\bf \[[S III] $\lambda$9532} & -0.632 & 52.4 & & 48.0 & $\pm$10.47 & 50.3 & 29.3\\
P7 $\lambda$9546 & -0.633 & 2.95 & & 2.71 & $\pm$0.59 & 3.49 & 3.61\\
HI 9-6 $\lambda$5.91$\mu$m & -0.988 & 0.410 & & 0.358 & $\pm$0.123 & 0.071 & 0.076\\
\[[Ar II] $\lambda$6.99$\mu$m & -0.990 & 0.770 & & 0.672 & $\pm$0.232 & 0.288 & 0.532\\
HI 6-5 $\lambda$7.45$\mu$m & -0.990 & 4.48 & & 3.91 & $\pm$1.35 & 0.148 & 0.160\\
\[[Ar III] $\lambda$8.99$\mu$m & -0.959 & 5.25 & & 4.60 & $\pm$1.53 & 5.41 & 4.82\\
{\bf \[[S IV] $\lambda$10.52$\mu$m} & -0.959 & 10.1 & & 8.86 & $\pm$2.95 & 7.59 & 3.52\\
HI 9-7 $\lambda$11.31$\mu$m & -0.970 & 0.338 & & 0.296 & $\pm$0.100 & 0.269 & 0.291\\
HI 7-6 $\lambda$12.37$\mu$m & -0.980 & 0.996 & & 0.871 & $\pm$0.297 & 0.839 & 0.904\\
\[[Ne II] $\lambda$12.80$\mu$m & -0.983 & 7.85 & & 6.86 & $\pm$2.35 & 1.72 & 5.24\\
\[[Ne III] $\lambda$15.50$\mu$m & -0.985 & 29.6 & & 25.9 & $\pm$8.87 & 16.7 & 14.7\\
\[[S III] $\lambda$18.70$\mu$m & -0.981 & 13.3 & & 11.6 & $\pm$3.97 & 17.5 & 11.5\\
\[[Fe III] $\lambda$22.93$\mu$m & -0.987 & 1.84 & & 1.61 & $\pm$0.55 & 1.34 & 1.53\\
\[[S III] $\lambda$33.48$\mu$m & -0.993 & 5.74 & & 5.01 & $\pm$1.73 & 6.13 & 3.81\\
\[[Si II] $\lambda$34.81$\mu$m & -0.993 & 0.700 & & 0.611 & $\pm$0.211 & 0.762 & 0.229\\
\[[Ne III] $\lambda$36.01$\mu$m & -0.993 & 3.50 & & 3.05 & $\pm$1.06 & 1.39 & 1.21\\
c & \nodata & 0.06 &&\nodata&&\nodata&\nodata \\
H$\alpha$/H$\beta$ &\nodata & 2.82 &&\nodata&&\nodata&\nodata \\
log F$_{H\beta}$\tablenotemark{b} & \nodata  & -11.78 & & \nodata & &-11.63\tablenotemark{c} & -9.71\tablenotemark{c} \\
\enddata
\tablenotetext{a}{Deblended.}
\tablenotetext{b}{ergs\ cm$^{-2}$ s$^{-1}$ in our extracted spectra}
\tablenotetext{c}{A distance of 11.4~kpc from \citet{CKS92} was used to obtain this value from the H$\beta$ luminosity predicted by the model.}
\end{deluxetable}

\newpage

\begin{deluxetable}{lcc}
\tabletypesize{\normalsize}
\setlength{\tabcolsep}{0.07in}
\tablecolumns{3}
\tablewidth{0in}
\tablecaption{Spitzer Telescope Observations\label{coverage}}
\tablehead{
\colhead{Module} & \colhead{Wavelength Range ($\mu$m)} & \colhead{Integration Time(s)} 
}
\startdata
SL2 & 5.2-8.7 & 960 \\
SL1 & 7.4-14.5 & 240 \\
SH & 9.9-19.6 & 240 \\
LH & 18.7-37.2 & 3600
\enddata
\end{deluxetable}

\newpage

\begin{deluxetable}{lccc}
\tabletypesize{\normalsize}
\setlength{\tabcolsep}{0.07in}
\tablecolumns{4}
\tablewidth{0in}
\tablecaption{Radial Velocity of DdDm-1\label{radvel}}
\tablehead{
\colhead{Ion} & \colhead{Observed Wavelength} & \colhead{Rest Wavelength} & \colhead{Radial Velocity (km~s$^{-1}$)}
}

\startdata

\[[O III]	 &			4358.81	&	4363.23	&	-303.9 \\
H$\gamma$ &		4336.12 &		4340.47	&	-300.7\\
He I	&			4467.01 &		4471.5 &		-301.2\\
H$\beta$ &		4856.45 &		4861.33 & 	-301.2\\
\[[O III] 	&		4953.95 &		4958.91 &		-300.1\\
\[[O III]	&			5001.84 &		5006.84 &		-299.6\\
\[[N II]	&			5748.84 &		5754.6 &		-300.3\\
He I	&			5869.76	&	5875.66 &		-301.2\\
\[[N II]	&			6541.51	&	6548.1  &		-301.9\\
H$\alpha$ 	&	6556.21	&	6562.77  &	-299.9\\
\[[N II]	&			6576.85 &		6583.5 &		-303.0\\
\[[S II]	&			6709.75 &		6716.44 &		-298.8\\
\[[S II] &			6724.09 &		6730.82	&	-300.0\\
He I &			7058.15 &		7065.25 &		-301.5\\
average measured &\nodata&\nodata&  -301$\pm$1.4\\
correction to heliocentric &\nodata&\nodata& -9.4\\
{\bf final radial velocity} &\nodata&\nodata& -310.3 $\pm$1.4\\
\enddata
\end{deluxetable}

\newpage

\begin{deluxetable}{lccccc}
\tabletypesize{\small}
\setlength{\tabcolsep}{0.07in}
\tablecolumns{6}
\tablewidth{0in}
\tablecaption{Temperatures and Densities\label{td}}
\tablehead{
\colhead{Parameter\tablenotemark{a}} &
\colhead{This Paper} &
\colhead{C87\tablenotemark{b}} &
\colhead{BC84\tablenotemark{c}} &
\colhead{HBK04} &
\colhead{WLB05\tablenotemark{d}}
}
\startdata
T$_{[O III]}$ & 12060($\pm$690) & 11800($\pm$800) & 12100($\pm$600)  & 11700 & 12300\\
T$_{[N II]}$ & 12940$\pm$1236) & 11000($\pm$1200) & 12800($\pm$1000) & 11400 & 12980 \\
T$_{[O II]}$ & 15830($\pm$9500) & \nodata &11000($\pm$2000) & 10100 & 16110\\
T$_{[S II]}$ & 15650($\pm$13760)\tablenotemark{e} & \nodata & 18000($\pm$6000) & 7900 & 13850\\
T$_{[S III]}$ & 12950($\pm$1392)\tablenotemark{f} & \nodata & 11500($\pm$3000) & 12700 & \nodata \\
N$_e$$_{[S II]}$ & 4092($\pm$2491) & 4400 & 3200($\pm$1300) & 4000 & \nodata \\
N$_e$$_{[Cl III]}$ & 3973($\pm$4427) & \nodata & \nodata & \nodata & \nodata \\
N$_e$$_{[S III]}$ & 3293($\pm$743) & \nodata & \nodata & \nodata & \nodata \\
\enddata
\tablenotetext{a}{Wavelengths (in {\AA}ngstroms, unless noted otherwise) of emission lines used to calculate electron temperatures and densities were: T([O~III]): 5007, 4363; T([N~II]): 6584, 5755; T([O~II]): 3727, 7324; T([S~II]): 6716, 6731, 4071; T([S~III]): 9532, 9069, 6312; N$_e$([S~II]): 6716, 6731; N$_e$([Cl~III]): 5518, 5538; N$_e$([S~III]): 18.7$\mu$m, 33.5$\mu$m. Temperatures and densities are expressed in Kelvins and cm$^{-3}$, respectively.}
\tablenotetext{b}{\citet{C87}}
\tablenotetext{c}{\citet{BC84}}
\tablenotetext{d}{\citet{WLB05}}
\tablenotetext{e}{We note that the value of 7900 K for T$_{[SII]}$ was misreported in \citet{HKB04}.}
\tablenotetext{f}{This temperature was based on the $\lambda$9532 line only, since the $\lambda$9532/$\lambda$9069 ratio exceeded the theoretical value, indicating that $\lambda$9069 is partially absorbed by the Earth's atmosphere.}
\end{deluxetable}

\newpage

\begin{deluxetable}{lcccc}
\tabletypesize{\small}
\setlength{\tabcolsep}{0.07in}
\tablecolumns{5}
\tablewidth{0in}
\tablecaption{Ionic Abundances\label{ions}}
\tablehead{
\colhead{Ion} &
\colhead{T$_{\mathrm{used}}$} &
\colhead{This Paper} &
\colhead{C87\tablenotemark{a}} &
\colhead{BC84\tablenotemark{b}} 
}
\startdata
He$^{+}$ & [O III] & 9.34$\pm$1.28(-2) & 0.11 & 0.084 \\
He$^{+2}$ & [O III] & 3.53$\pm$2.14(-4) & $<$5.0(-4) & $<$3.0(-4) \\
icf(He) &  & 1.00 & & \\
\\
O$^{0}$(6300) & [N II] & \tablenotemark{*}2.05$\pm$0.60(-6) & 4.6(-6) & \nodata \\
O$^{0}$(6363) & [N II] &  \tablenotemark{*}2.23$\pm$0.65(-6) & \nodata & \nodata \\
O$^{0}$ & wm & \tablenotemark{\dag}2.10$\pm$0.60(-6) & \nodata & \nodata \\
O$^{+}$(3727) & [N II] &  \tablenotemark{*}2.64$\pm$1.10(-5) & 4.7(-5) & 3.0(-5) \\
O$^{+}$(7325) & [N II] &  \tablenotemark{*}3.23$\pm$1.36(-5) & \nodata & \nodata \\
O$^{+}$ & wm & 2.72$\pm$1.00(-5) & \nodata& \nodata \\
O$^{+2}$(5007) & [O III] &  \tablenotemark{*}8.88$\pm$2.27(-5) & 8.96(-5) & 7.6(-5)\\
O$^{+2}$(4959) & [O III] &  \tablenotemark{*}8.41$\pm$1.72(-5) & \nodata & \nodata \\
O$^{+2}$(4363) & [O III] &  \tablenotemark{*}8.88$\pm$2.27(-5) & \nodata & \nodata \\
O$^{+2}$ & wm & 8.76$\pm$2.08(-5) & \nodata & \nodata \\
icf(O) &  & 1.00 & 1.0\tablenotemark{c} & 1.0 \\
\\
N$^{+}$(6584) & [N II] &  \tablenotemark{*}6.10$\pm$1.61(-6) & 8.00(-6) & 5.6(-6) \\
N$^{+}$(6548) & [N II] &  \tablenotemark{*}5.65$\pm$1.33(-6) & \nodata & \nodata \\
N$^{+}$(5755) & [N II] &  \tablenotemark{*}6.10$\pm$1.61(-6) & \nodata & \nodata \\
N$^{+}$ & wm & 5.99$\pm$1.51(-6) & \nodata & \nodata \\
N$^{+2}$(1751) & [O III] & 4.08$\pm$2.92(-6) & \nodata & \nodata \\
icf(N) &  & 4.22 & 2.95 & 3.53 \\
\\
C$^{+2}$(1909) & [O III] & 8.53$\pm$4.63(-6) & $<$6.9(-6) & 4.5(-4)\tablenotemark{d}\\
icf(C) & & 1.31 & & 1.5\\
\\
Ne$^{+}$(12.8$\mu m$) & [O III] &  \tablenotemark{*}8.46$\pm$3.05(-6) & \nodata & \nodata \\
Ne$^{+2}$(3869) & [O III] &  \tablenotemark{*}1.54$\pm$0.37(-5) & 1.3(-5) & 1.5(-5) \\
Ne$^{+2}$(3967) & [O III] & 1.96$\pm$0.89(-5) & \nodata & \nodata \\
Ne$^{+2}$(15.5$\mu m$) & [O III] &  \tablenotemark{*}1.58$\pm$0.57(-5) & \nodata & \nodata \\
Ne$^{+2}$(36.0$\mu m$) & [O III] &  \tablenotemark{*}2.28$\pm$0.84(-5) & \nodata & \nodata \\
Ne$^{+2}$ & wm & 1.60$\pm$0.33(-5) & \nodata & \nodata\\
icf(Ne) &  & 1.31 & 1.5 & 1.39 \\
\\
S$^{+}$ (6717+6731)& [N II] &  \tablenotemark{*}1.83$\pm$0.71(-7) & 2.7(-7) & 1.6(-7) \\
S$^{+}$(6716) & [N II] & 1.83$\pm$0.71(-7) & \nodata & \nodata \\
S$^{+}$(6731) & [N II] & 1.83$\pm$0.70(-7) & \nodata & \nodata \\
S$^{+}$ & wm & 1.50$\pm$1.64(-7) & \nodata & \nodata \\
S$^{+2}$(9532) & [S III] &  \tablenotemark{*}1.73$\pm$0.55(-6) & \nodata & \nodata\\
S$^{+2}$(6312) & [S III] &  \tablenotemark{*}1.73$\pm$0.55(-6) & 2.56(-6) & 2.3(-6) \\
S$^{+2}$(18.7$\mu m$) & [S III] &  \tablenotemark{*}1.38$\pm$0.55(-6) & \nodata & \nodata \\
S$^{+2}$(33.4$\mu m$) & [S III] &  \tablenotemark{*}1.59$\pm$1.05(-6) & \nodata & \nodata \\
S$^{+2}$ & wm & 1.66$\pm$0.54(-6) & \nodata & \nodata \\
S$^{+3}$(10.5$\mu m$) & [O III] &  \tablenotemark{*}2.10$\pm$0.78(-7)  & \nodata & \nodata \\
icf(S) &  & 1.17 & 1.2 & 1.17 \\
\\
Ar$^{+}$(7.0$\mu m$) & [N II] &  \tablenotemark{*}5.28$\pm$1.93(-8) & \nodata & \nodata \\
Ar$^{+2}$(7135) & [O III] &  \tablenotemark{*}4.72$\pm$1.11(-7) & 4.9(-7) & 3.2(-7) \\
Ar$^{+2}$(7751) & [O III] &  \tablenotemark{*}4.66$\pm$1.21(-7) & \nodata & \nodata \\
Ar$^{+2}$(9.0$\mu m$) & [O III] &  \tablenotemark{*}4.36$\pm$1.54(-7) & \nodata & \nodata \\
Ar$^{+2}$ & wm & 4.59$\pm$1.19(-7) & \nodata & \nodata \\
Ar$^{+3}$(4740) & [O III] &  \tablenotemark{*}1.85$\pm$0.36(-8) & $<$4.0(-8) & \nodata \\
Ar$^{+4}$(7005) & [O III] &  \tablenotemark{*}1.18$\pm$0.29(-8) & \nodata & \nodata \\
icf(Ar) &  & 1.31 & 1.5 & 1.5\tablenotemark{e}\\
\\
Cl$^{+}$(8578) & [N II] &  \tablenotemark{*}8.62$\pm$2.40(-9) & \nodata & \nodata \\
Cl$^{+2}$(5517) & [S III] &  \tablenotemark{*}1.81$\pm$0.75(-8) & \nodata & \nodata \\
Cl$^{+2}$(5537) & [S III] &  \tablenotemark{*}1.79$\pm$0.64(-8) & \nodata & \nodata \\
Cl$^{+2}$ & wm & 1.80$\pm$0.54(-8) & \nodata & \nodata \\
icf(Cl) &  & 1.00 & \nodata & \nodata \\
\enddata
\tablenotetext{a}{\citet{C87}}
\tablenotetext{b}{\citet{BC84}}
\tablenotetext{c}{\citet{C87} included the O$^o$, O$^+$, and O$^{+2}$ abundances and assumed an ICF of unity}
\tablenotetext{d}{Abundance derived from C II $\lambda$4267}
\tablenotetext{e}{Based on Ar$^{+2}$ only}
\tablenotetext{*}{value included in weighted mean (wm), \\weighted by observed flux}\\
\tablenotetext{\dag}{not included in total oxygen abundance calculation}
\end{deluxetable}

\newpage

\begin{deluxetable}{lc}
\tabletypesize{\small}
\setlength{\tabcolsep}{0.07in}
\tablecolumns{2}
\tablewidth{0in}
\tablecaption{Empirical Elemental Abundances\label{elements}}
\tablehead{
\colhead{Element} &
\colhead{12+log(X/H)}
}
\startdata
He/H & 10.97 \\
O/H & 8.06 \\
N/H & 7.40 \\
C/H & 7.05 \\
Ne/H & 7.32 \\
S/H & 6.33 \\
S(w/[S IV])/H & 6.31 \\
Ar/H & 5.81 \\
Cl/H & 4.42 \\
\enddata
\end{deluxetable}

\clearpage

 \begin{deluxetable}{lcc}
 \tabletypesize{\normalsize}
 \setlength{\tabcolsep}{0.07in}
 \tablecolumns{3}
 \tablewidth{0in}
 \tablecaption{Model Parameters\label{parameters}}
 \tablehead{
 \colhead{Parameter\tablenotemark{a}} &
 \colhead{Model 18} &
 \colhead{Model 32}
} 
 \startdata
He/H  & 11.0 & 11.0 \\
C/H & 6.55 & 6.85 \\
N/H & 7.56 & 7.56 \\
O/H & 8.06 & 8.06 \\
Ne/H & 7.16 & 7.30 \\
S/H & 6.45 & 6.25 \\
Cl/H & 4.25 & 4.25 \\
Ar/H & 5.81 & 5.81 \\
Fe/H & 6.11 & 6.11 \\
 $T_{eff}$ (K) & 55,000\tablenotemark{b} & 40,000\tablenotemark{c} \\
 log L/L$_{\odot}$ (ergs s$^{-1}$) & 3.0 & 5.0 \\
 Total Density (cm$^{-3}$) & 3980 & 3980 \\
 Radius (pc) & 0.032 & 0.032 \\
 Filling Factor & 1 & 0.5 \\
 \enddata
\tablenotetext{a}{Abundances are expressed in the format 12+log(X/H)}
 \tablenotetext{b}{Central star model atmosphere from Rauch (2002); log g=6.5, Z=0.1Z$_{\odot}$}
 \tablenotetext{c}{Blackbody}
 \end{deluxetable}

\newpage

\begin{deluxetable}{lcccccccc}
\tabletypesize{\normalsize}
\setlength{\tabcolsep}{0.07in}
\tablecolumns{9}
\tablewidth{0in}
\tablecaption{Final Abundances\tablenotemark{a}\label{final}}
\tablehead{
\colhead{Element Ratio} &
\colhead{Correction Factor} &
\colhead{This Paper} &
\colhead{C87\tablenotemark{b}} &
\colhead{BC84\tablenotemark{c}} &
\colhead{WLB05\tablenotemark{d}} &
\colhead{Sun\tablenotemark{e}} &
\colhead{Orion\tablenotemark{f}} &
\colhead{[X]\tablenotemark{g}}
}
\startdata
He/H & 1.16 & 11.03$\pm$.056 & 11.00 & 11.02 & 10.95 & 10.93 & 10.99 & +0.10 \\
C/H & \nodata & 6.55\tablenotemark{h}$\pm$.20 & 7.14 & 8.83  & 6.91 & 8.39 & 8.39 & -1.84 \\
N/H & 1.64 & 7.62$\pm$.11 & 7.40 & 7.30 & 7.31 & 7.78 & 7.78 & -0.16 \\
O/H & 1.40 & 8.20$\pm$.085 & 8.15 & 8.04 & 8.05 & 8.66 & 8.63 & -0.46 \\
Ne/H & 1.13 & 7.37$\pm$.080 & 7.30 & 7.32 & 7.24 & 7.84 & 7.89 & -0.47 \\
Si/H & \nodata & 6.15\tablenotemark{h}$\pm$.13 & 6.30 & \nodata & \nodata & 7.51 & \nodata & -1.36 \\
S/H & 1.34 & 6.46$\pm$.12 & 6.46 & 6.46 & 6.35 & 7.14 & 7.17 & -0.68 \\
Cl/H & 1.23 & 4.51$\pm$.10 & \nodata & \nodata & 4.69 & 5.50 & 5.33 & -0.99 \\
Ar/H & 1.32 & 5.93$\pm$.096 & \nodata & 5.68 & 5.16 & 6.18 & 6.80 & -0.25 \\
Fe/H & \nodata & 6.10\tablenotemark{h}$\pm$.053 & 6.32 & \nodata & \nodata & 7.45 & 6.41 & -1.35 \\
C/O & \nodata & -1.65$\pm$.20 & -1.01 & +0.79 & -1.14 & -0.27 & -0.24 & -1.38 \\
N/O & \nodata & -0.58$\pm$.10 & -0.75 & -0.74 & -0.74 & -0.88 & -0.85 & +0.30 \\
O/Fe & \nodata & +2.1$\pm$.14 & +1.83 & \nodata & \nodata & +1.21 & +2.22 & +0.89 \\
\enddata
\tablenotetext{a}{Elemental abundances of X/H in columns 3-7 are expressed in the format 12+log(X), where X is the element ratio in column 1. Heavy element ratios X/Y in the last three rows are expressed as log(X/Y).}
\tablenotetext{b}{\citet{C87}; final abundances from their Table~9}
\tablenotetext{c}{\citet{BC84}}
\tablenotetext{d}{\citet{WLB05}}
\tablenotetext{e}{\citet{AGS05}}
\tablenotetext{f}{\citet{E98}; gas phase abundances}
\tablenotetext{g}{[X]=log(X)-log(X)$_{\odot}$, where X is the element ratio in column 1}
\tablenotetext{h}{Based upon the model value only}
\end{deluxetable}

\newpage

 \begin{deluxetable}{lcc}
 \tabletypesize{\normalsize}
 \setlength{\tabcolsep}{0.07in}
 \tablecolumns{3}
 \tablewidth{0in}
 \tablecaption{Comparison with \citet{D03}\label{dinerstein}}
 \tablehead{
 \colhead{Parameter} &
 \colhead{Dinerstein et al.} &
 \colhead{This Paper}
} 
 \startdata
S$^+$/H$^+$ & $2.2 \times 10^{-7}$ & $1.5 \times 10^{-7}$ \\
S$^{+2}$/H$^+$ & $1.63 \times 10^{-6}$ & $1.66 \times 10^{-6}$ \\
S$^{+3}$/H$^+$ & $2.3 \times 10^{-7}$ & $2.1 \times 10^{-7}$ \\
Total S/H & $2.1 \times 10^{-6}$ & $2.9 \times 10^{-6}$ \\
Ne$^+$/H$^+$ & $2.1 \times 10^{-6}$ & $8.46 \times 10^{-6}$ \\
Ne$^{+2}$/H$^+$ & $1.17 \times 10^{-5}$ & $1.6 \times 10^{-5}$ \\
Total Ne/H & $1.4 \times 10^{-5}$ & $2.3 \times 10^{-5}$
 \enddata
 \end{deluxetable}

\clearpage

\end{document}